\documentclass[11pt, usenatbib]{gh2013}

\usepackage{mathptmx}       
\usepackage{helvet}         
\usepackage{courier}        
\usepackage{makeidx}         
\usepackage{graphicx}        
\usepackage{multicol}        
\usepackage{natbib} 

\begin{document}

\title*{The impact of metallicity and dynamics on the evolution of young star clusters}
\titlerunning{The evolution of young star clusters}
\author{Michela Mapelli$^{1}$, Alessandro Alberto Trani$^{1,2}$ and Alessandro Bressan$^{1,2}$}\authorrunning{Mapelli, Trani \&{} Bressan}
\institute{$^{1}$ INAF, Osservatorio Astronomico di Padova, Vicolo dell'Osservatorio 5, I--35122, Padova, Italy, \email{michela.mapelli@oapd.inaf.it}
\\ $^{2}$ SISSA/ISAS, via Bonomea 265, I--34136, Trieste, Italy }
%
%
\maketitle


\vskip -3.5 cm  
\abstract{
The early evolution of a dense young star cluster (YSC) depends on the intricate connection between stellar evolution and dynamical processes. Thus, N-body simulations of YSCs must account for both aspects. We discuss N-body simulations of YSCs with three different metallicities ($Z=0.01$, 0.1 and 1 Z$_\odot$), including metallicity-dependent stellar evolution recipes and metallicity-dependent prescriptions for stellar winds and remnant formation.  We show that mass-loss by stellar winds influences the reversal of core collapse. 
In particular, the post-collapse expansion of the core is faster in metal-rich YSCs than in metal-poor YSCs, because the former lose more mass (through stellar winds) than the latter. As a consequence, the half-mass radius expands more in metal-poor YSCs. 
We also discuss how these findings depend on the total mass and on the virial radius of the YSC. These results give us a clue to understand the early evolution of YSCs with different metallicity.
}
\section{Introduction}
\label{sec:Introduction}
The densest young star clusters (YSCs) are collisional environments: their central two-body relaxation timescale ($t_{\rm rlx}$) is generally shorter than $100$ Myr. For a realistic initial mass function (IMF), YSCs are expected to undergo core collapse (CC) in less than $t_{\rm rlx}$. 
The reversal of CC is generally ascribed to hard binaries (i.e. binaries whose binding energy is higher than the average kinetic energy of a star in the cluster), because they transfer kinetic energy to stars through three-body encounters.


It has long been debated whether mass-loss by stellar winds and/or supernovae (SNe) is efficient in affecting CC (e.g. \citealt{mapelli&bressan2013}, and references therein). 
Stellar winds and SNe eject mass from a star cluster, making the central potential well shallower and quenching the onset of gravothermal instability. Furthermore, stellar winds are expected to depend on  metallicity ($Z$): metal-poor stars lose less mass than metal-rich ones (e.g. \citealt{vink2001}). Thus, the effect of stellar winds on CC is expected to be stronger at high $Z$. In this paper, we investigate the impact of $Z$-dependent stellar winds and SNe on the CC of YSCs, by means of direct-summation N-body simulations.
\section{Method and simulations}
\label{sec:Simulations}
We adopt 
the {\sc starlab} public software environment (\citealt{portegies2001}), in the version modified by \cite{mapelli2013}. This version includes recipes for $Z$-dependent stellar evolution (\citealt{hurley2000}), stellar winds (\citealt{vink2001}) and direct collapse of massive metal-poor stars (\citealt{mapelli2009}).  
The initial conditions of the simulated YSCs have been generated following a King profile. 
The mass of each simulated particle (which corresponds to a single star) has been randomly drawn from a Kroupa IMF with minimum and maximum mass 0.1 and 150 M$_\odot$, respectively. 
The runs discussed in this paper are the following.

{\bf Set A:} 300 N-body realizations of YSCs with virial radius $r_{\rm vir}=1$ pc, dimensionless central potential $W_0=5$, particle number $N_\ast{}=5\times{}10^3$. These runs have already been discussed in \cite{mapelli&bressan2013}.

{\bf Set B:} 30 N-body realizations of YSCs with $r_{\rm vir}=1$ pc, $W_0=5$, $N_\ast{}=5\times{}10^4$.  

{\bf Set C:} 30 N-body realizations of YSCs with $r_{\rm vir}=5$ pc, $W_0=5$, $N_\ast{}=5\times{}10^4$.  

In each set of simulations,  $1/3$ of the runs have $Z=$ Z$_\odot$, $1/3$ have $Z=0.1$ Z$_\odot$ and $1/3$ have $Z=0.01$ Z$_\odot$. The structural parameters (core radius $r_{\rm c}$, half-mass radius $r_{\rm hm}$, half-light radius  $r_{\rm hl}$ and total binary binding energy $E_{\rm b}$) discussed in the following are the median value of different N-body realizations (with the same $Z$) in each set of runs, to filter out statistical fluctuations.
\section{Results}
\label{sec:Results}
\begin{figure*}
  \centering
  \includegraphics[width=  11.5 cm]{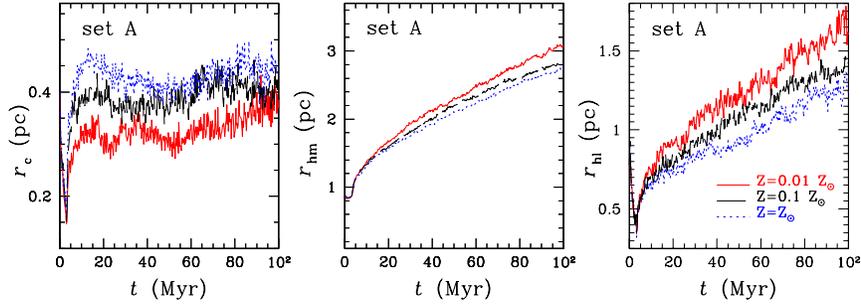}
  \caption{Core radius $r_{\rm c}$ (left-hand panel), half-mass radius $r_{\rm hm}$ (central panel) and half-light radius $r_{\rm hl}$ (right-hand panel) as a function of time for $Z=0.01$ Z$_\odot$ (red solid line), 0.1 Z$_\odot$ (black dashed line) and 1 Z$_\odot$ (blue dotted line). Runs of set A are shown ($W_0=5$, $r_{\rm vir}=1$ pc, $N_{\ast{}}=5\times{}10^3$). Each line is the median value of 100 N-body realizations.}
  \label{fig:fig1}
\end{figure*}
Fig.~\ref{fig:fig1} shows the behaviour of $r_{\rm c}$, $r_{\rm hm}$ and  $r_{\rm hl}$ in the runs of set~A, for different $Z$. The YSCs in set~A  undergo CC at $t\sim{}2-3$ Myr (regardless of $Z$). 
The impact of $Z$-dependent stellar winds and SNe is apparent in the reversal of CC:  $r_{\rm c}$  increases faster in metal-rich YSCs, because stellar winds are stronger than in metal-poor YSCs. At the same time, $r_{\rm hm}$ expands more in metal-poor YSCs than in metal-rich YSCs. The reason is that three-body encounters are more efficient in metal-poor YSCs (where higher core densities are reached during the CC) and inject more kinetic energy in the halo. The same trend is apparent in the half-light radius $r_{\rm hl}$: it expands more in metal-poor YSCs than in metal-rich ones. Furthermore, the difference between the half-light radii of metal-rich and metal-poor YSCs is a factor of two larger than the difference between the half-mass radii, due to mass segregation and to the $Z$-dependence of the stellar luminosity (\citealt{mapelli&bressan2013}). 

\begin{figure*}
  \centering
  \includegraphics[width=  10.0 cm]{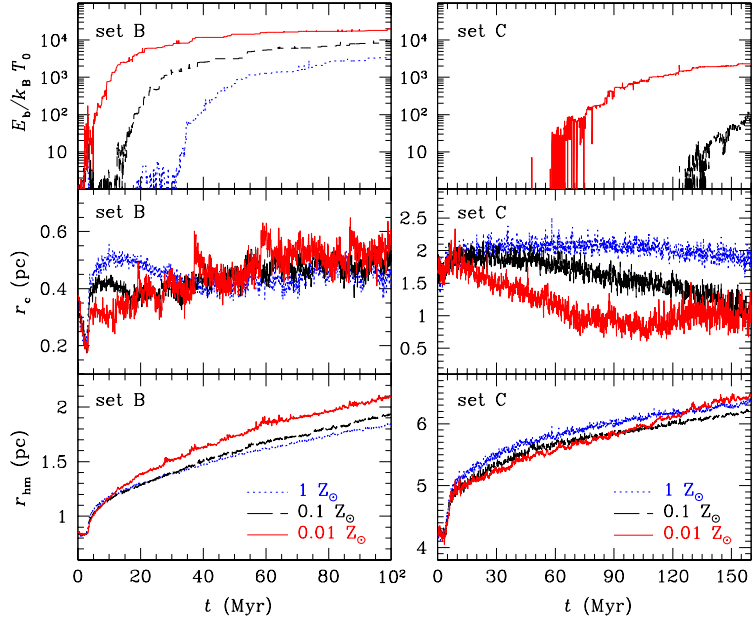}
  \caption{Total binary binding energy $E_{\rm b}$ (top panel), core radius $r_{\rm c}$ (central panel) and half-mass radius $r_{\rm hm}$ (bottom panel) as a function of time for $Z=0.01$ Z$_\odot$ (red solid line), 0.1 Z$_\odot$ (black dashed line) and 1 Z$_\odot$ (blue dotted line).  $E_{\rm b}$ is normalized to the average initial kinetic energy of a star in the YSC ($k_{\rm B}\,{}T_0$). Each line is the median value of 10 runs. Left-hand panel: set B ($W_0=5$, $r_{\rm vir}=1$ pc, $N_{\ast{}}=5\times{}10^4$). Right-hand panel: set C ($W_0=5$, $r_{\rm vir}=5$ pc, $N_{\ast{}}=5\times{}10^4$).}
  \label{fig:fig2}
\end{figure*}

How much do these results 
depend on the mass and on the size of YSCs? The left-hand column of Fig.~\ref{fig:fig2} shows the behaviour of $r_{\rm c}$ and $r_{\rm hm}$ in runs of set~B, which are 10 times more massive than runs of set~A, even if they have the same size ($r_{\rm vir}=1$ pc). The CC occurs again at $t\sim{}2-3$ Myr.  $r_{\rm c}$ and $r_{\rm hm}$ show the same trend in set~B and in set~A. 
The right-hand column of Fig.~\ref{fig:fig2} shows the behaviour of $r_{\rm c}$ and $r_{\rm hm}$ in runs of set~C, which are as massive as runs of set~B but have larger  size ($r_{\rm vir}=5$ pc). Thus, the YSCs of set~C have lower central density, and the CC is expected to occur at later times ($t_{\rm rlx}\propto{}r_{\rm hm}^{3/2}$). In fact, the CC begins at $\sim{}60$ Myr for $Z=0.01$ Z$_\odot$ and at later times for higher $Z$. In the metal-rich ($Z=0.1, 1$ Z$_\odot$) YSCs of set~C the CC is delayed, because the mass-loss by stellar winds is sufficient to keep the core stable against collapse. This implies that $r_{\rm hm}$ is marginally larger in metal-rich YSCs than in metal-poor ones, until the CC begins. During the CC, the half-mass radius of metal-poor YSCs starts expanding faster than that of metal-rich YSCs, producing the same trend as observed in set~A and B.

This interpretation is confirmed by the top row of Fig.~\ref{fig:fig2}, which shows the total binding energy of binaries $E_{\rm b}$ (considering all binaries in a YSC at a given time). Since we do not include primordial binaries in our simulations, $E_{\rm b}$ represents the total energy that is stored in binaries as a consequence of three-body encounters. The YSCs of set~C have $E_{\rm b}\sim{}0$ up to $t\sim{}60$ Myr. In contrast, $E_{\rm b}$ in runs of set~B grows dramatically during the first CC, decreases during the rapid expansion phase and grows steadily at later times. Thus, three-body encounters are almost negligible in the loose YSCs of set~C, while they are the main engine of CC reversal in sets~A and B.
%



\section{Conclusions}
We have shown that stellar winds are very important in the early evolution of YSCs and that their effects strongly depend on $Z$. In particular, the post-collapse re-expansion of the core is faster for metal-rich YSCs than for metal-poor YSCs, because the former lose more mass (through stellar winds) than the latter. As a consequence, the half-mass radius and the half-light radius expand faster in metal-poor YSCs. 
The initial size of the YSC plays a critical role, because the relaxation timescale and thus the onset of CC strongly depend on it ($t_{\rm rlx}\propto{}r_{\rm hm}^{3/2}$). The total mass of the YSC has only marginal effects. Other YSC properties (e.g. $W_0$) deserve further investigation. Furthermore, the $Z$-dependence of stellar winds is still barely understood, especially in the post-main sequence evolution. Thus, forthcoming studies will also investigate the effects of different recipes of mass-loss by stellar winds.

\begin{acknowledgement}
MM acknowledges financial support from MIUR through grant FIRB 2012 RBFR12PM1F, from INAF through grant PRIN-2011-1 and from CONACyT through grant 169554.  
The authors thank the organizers and the participants of the fifth Guillermo Haro conference for the excellent organization and for the stimulating discussions.
\end{acknowledgement}

\end{document}